\documentclass[9pt,twocolumn,letter]{article}

\usepackage[left=0.5in,right=0.5in,top=0.5in,bottom=1in]{geometry}  
\usepackage{graphics}
\usepackage{graphicx}
\usepackage[title]{appendix}
\usepackage[colorlinks = true,
            linkcolor = blue,
            urlcolor  = blue,
            citecolor = blue,
            anchorcolor = blue]{hyperref}
\usepackage{multibib}

\usepackage{mathtools,stmaryrd,soul}
\usepackage{amsmath,amssymb}
\DeclareMathOperator\Arg{Arg}
\usepackage{authblk}
\usepackage{xcolor}
\usepackage{bm}

\usepackage[utf8]{inputenc}

\usepackage{caption}

\newcommand{\beginsupplement}{%
        \setcounter{table}{0}
        \renewcommand{\thetable}{S\arabic{table}}%
        \setcounter{figure}{0}
        \renewcommand{\thefigure}{S\arabic{figure}}%
        \setcounter{section}{0}
        \renewcommand{\thesection}{S\arabic{section}}
     }

\usepackage{tikz}
\usetikzlibrary{shapes.geometric, arrows}
\tikzstyle{startstop} = [rectangle, rounded corners, minimum width=4cm, minimum height=1cm,text centered, draw=black, fill=red!30]
\tikzstyle{process} = [rectangle, minimum width=3cm, minimum height=1cm, text centered, draw=black, align=center, text width = 5cm]
\tikzstyle{decision} = [diamond, minimum width=3cm, minimum height=1cm, text centered, draw=black, fill=gray!30, align=center, text width = 2.2cm, aspect=1.3]
\tikzstyle{arrow} = [thick,->,>=stealth]
\tikzstyle{io} = [trapezium, trapezium left angle=70, trapezium right angle=110, minimum width=3cm, minimum height=1.3cm, text centered, draw=black, fill=blue!30, align = center]

\date{}

\begin{document}

\title{A Polarimetric Wavefront Imager}

\author[1*]{Baptiste Blochet}
\author[1]{Gr{\'e}goire Lelu}
\author[2]{Miguel A. Alonso}
\author[1,3,4*]{Marc Guillon}

\affil[1]{Saints-Pères Paris Institute for the Neurosciences, CNRS UMR 8003, Université de Paris, 45 rue des Saints-Pères, Paris 75006, France}
\affil[2]{Aix Marseille Univ, CNRS, Centrale Med, Institut Fresnel, UMR 7249, 13397 Marseille, France}
\affil[3]{Institut Langevin, ESPCI Paris, Universit\'e PSL, CNRS, Paris 75005, France}
\affil[4]{Institut Universitaire de France, Paris, France}
\affil[*]{baptiste.blochet@u-paris.fr - marc.guillon@u-paris.fr}


\maketitle


\paragraph{Abstract.}

\textbf{
Imaging both the polarization and the wavefront of a light beam is a complex task that typically demands several intensity acquisitions. Furthermore, sequential acquisition solutions are incompatible with the monitoring of ultra-fast processes. As a possible solution for single-shot wavefront and full-Stokes polarimetric imaging, we propose here a vector-beam lateral shearing interferometer. The device, composed of a patterned polarization-modulating Hartmann mask placed in the close vicinity of a camera, encodes all the information in the fringe pattern of a single image acquisition. 
}

\paragraph{Introduction.}

While optical sensors can only measure intensity, an important part of the information is typically encoded in the spatial phase (or wavefront) of the beam, as well as in its polarization state (the relative amplitude and phase between two orthogonal polarization components). Imaging the wavefront and/or polarization is of key importance in many applications: -- Phase imaging is typically required for cophasing segmented lasers arrays~\cite{Primot_OL_16} or imaging thin transparent objects~\cite{Park_NP_18,Baffou_Arxiv_24} where it can provide a quantitative measure of the dry mass of cells~\cite{BARER1952}, -- Polarimetric imaging is of interest for quantitative biological and biomedical applications~\cite{Metha_NM_24, Booth_LSA_21}, but also for non-destructive testing \cite{Khadir_NC_21} as well as machine vision under specular reflections~\cite{Chang_NC_23} and through fog~\cite{Fade_AO_14} for hidden target detection~\cite{Goudail_JOSA_11}, -- Simultaneous phase and polarimetric imaging is of high importance for vectorial adaptive optics~\cite{Booth_VAO_23}, metrology of metasurfaces~\cite{Khadir_NC_21}, optical tomography~\cite{Song2023} and single-molecule localization microscopy~\cite{Bon2018a}. Phase and polarimetric imaging typically demand $3$ and $4$ intensity acquisitions at least, respectively. 
In total, $12$ images are thus recorded for a complete determination of the electromagnetic field in the optical range. Unfortunately, no simple instrument exists that can fully image electromagnetic wavefields in the optical frequency range in a single acquisition step. 

Many techniques have been developed to image either polarization or phase but far less solutions have been proposed to image both~\cite{Monneret_OE_15,Wegener_NC_18}. The mere combination of the independently developed phase and polarization imaging modalities faces major obstacles associated with cumulative technical complexity and price of the resulting optical systems~\cite{Deniak_AO_08}, and are additionally either sequential, incomplete in polarization~\cite{Monneret_OE_15}, or have low-resolution~\cite{Wegener_NC_18}.


Here, we propose an original method to image both the wavefront and the full polarization state of a light beam in a single acquisition step. The technique we propose is based on the same principles as lateral shearing interferometry (LSI)~\cite{Primot_AO_93,Primot_JOSA_95}, a simple reference-free achromatic and quantitative wavefront imaging technique. In LSI, a phase grating is used as a Hartmann mask and placed at a close distance from a camera sensor, so generating a dense grid of focii. LSI is thus the high-resolution analog of the highly popular Shack-Hartmann wavefront sensors. In LSI, intensity is encoded in the low spatial frequency content of the camera pattern while wavefront gradients are encoded as spatial-frequency modulations. 
Similarly, we suggest here using a polarization modulating Hartmann mask to encode the full-Stokes image as well as the wavefront of the beam. Our polarimetric approach can be categorized as a division of channeled-imaging technique~\cite{Shaw_AO_06, Deniak_AO_08, Oka_AO_11}, in contrast with division of aperture~\cite{Shaw_AO_06}, division of amplitude~\cite{Faraon_ACSPhot_18,Capasso_Science_19} and division of focal plane~\cite{Gottlieb_OE_21,Yao_LSA_23} solutions. The solution we propose for full-Stokes polarimetric wavefront imaging is compatible with partially coherent polarization states, is highly achromatic, provides high-resolution images ($\simeq 400\times 400~{\rm px^2}$), and is potentially compatible with a compact and monolithic design. 

\begin{figure}[h!]
\centering
\fbox{\includegraphics[width=0.95\linewidth]{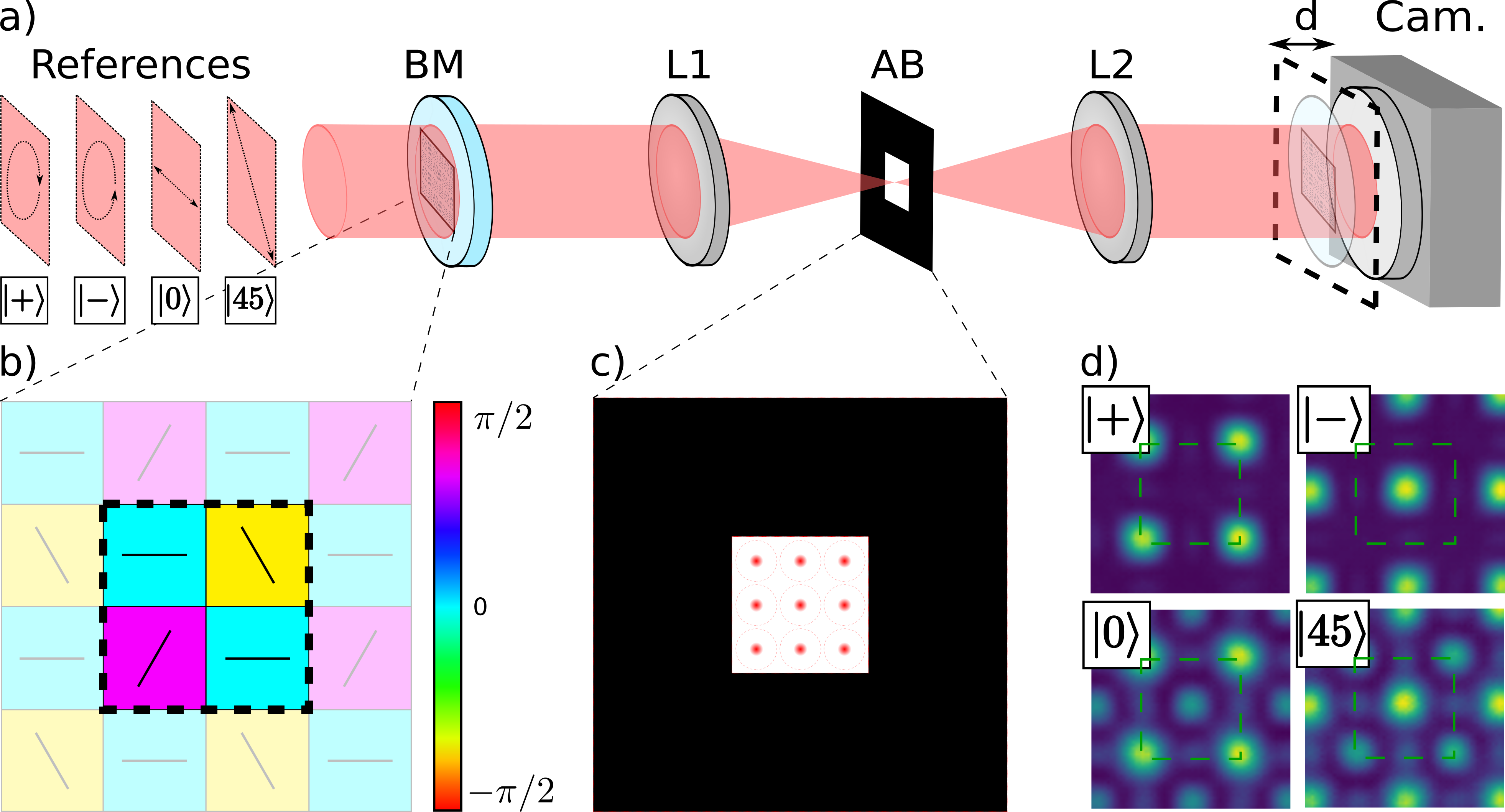}}
\caption{ {\bf Principle of the optical system.} 
(a) A periodic birefringent mask (BM) is imaged  at a distance $d$ from a camera sensor with a telescope. This polarization modulating diffractive mask 
whose unit cell is shown in (b), generates nine main diffracting orders (c) that can be optionally filtered with an aperture block (AB). (d) Specific intensity grids, forming a basis, are then obtained at the camera plane for impinging beams having polarization states defining a non-zero volume inside the Poincaré sphere. 
} 
\label{fig:principle}
\end{figure}

\paragraph{Principle of polarimetric LSI.}
We consider here a polarization-modulating Hartmann mask exhibiting linear birefringence. A linearly birefringent plate with slow and fast optical path differences (OPD) $\delta_s$ and $\delta_f$ changes an impinging circular polarization state $\bm{\sigma_\pm}$ to
\begin{align}
\bm{E}_{out,\pm}&= e^{i\phi}\left[ \cos\left(\frac{\varphi}{2}\right)\bm{\sigma_\pm} +i e^{\pm 2i\theta} \sin\left(\frac{\varphi}{2}\right)\bm{\sigma_\mp} \right],
\label{eq:bir-plate}
\end{align}
where $\phi = (\delta_s+\delta_f)/2$ is the mean phase delay, $\varphi = \delta_s-\delta_f$ the retardance, and $\theta$ the polar angle of the slow axis orientation (Supp.~\ref{app:linear_retarder_plate}). 
By patterning the orientation $\theta$ of the birefringent axis of the mask, the $e^{\pm 2i\theta}$ factor thus makes it possible to imprint a phase grating modulation as required by LSI. 
Note that the phase imprinted onto the converted cross-polarized circular beam is solely determined by the geometric patterning of the birefringence axis angle, so that the phase grating is perfectly achromatic. This principle is at the basis of numerous applications harnessing the Pancharatnam-Berry geometric phase in optics~\cite{pancharatnam1956, Berry1984}.
In addition, partial conversion of the $\bm{\sigma_\pm}$ components of the impinging beam to the cross-polarized states yields interferences accounting both for their respective amplitudes and relative phases. Placed at a close distance $d$ from a camera sensor, this mask can thus encode both the polarization state and the wavefront of the light beam into interference patterns (Fig.~\ref{fig:principle}a).
 
\paragraph{Design of the device.}
Let us consider a spatially coherent light beam characterized by a Stokes vector $\bm{s} = (s_0,s_1,s_2,s_3)$ and a local transverse wavevector $\bm{k_\perp}= k_x \bm{e_x}+k_y \bm{e_y}$. The intensity at the camera sensor placed at a distance $d$ from the mask is given by (Supp.~\ref{app:WFS_equation})
\begin{equation}
\bm{I}(\bm{r})=\bm{C}\left[\bm{r}+d\frac{\bm{k_\perp}(\bm{r})}{k_0}\right]\bm{s}(\bm{r}),
\label{eq:eq_WFS}
\end{equation}
where $\bm{r}=(x,y)$ is the transverse position vector, $k_0$ is the wavenumber of the light beam and $\bm{C} =(\bm{c_0},\bm{c_1},\bm{c_2},\bm{c_3})$ is a $4$-column matrix containing the $4$ respective normalized Stokes calibration intensity
patterns: $\bm{c_j}$ is the calibration intensity pattern representing the corresponding component of the Stokes vector $s_j$ only, in the case of a planar wavefront illumination. 
The wavefront map is derived from $\bm{k_\perp}(\bm{r})$ through a numerical two-dimensional integration.
Here, we find $s_j$ and $\bm{k_\perp}(\bm{r})$ by linearizing Eq.~\eqref{eq:eq_WFS}. For sufficiently small angles and distances $d$, the magnitude of the dimensionless displacement amount $\bm{\kappa} = \frac{d}{\Lambda k_0}\bm{k_\perp}$ is much smaller than $1$, where $\Lambda$ is the period of the intensity grid. A first-order Taylor expansion of $\bm{C}$ yields
\begin{equation}
\bm{I} \approx  \bm{C}\bm{s} + \kappa_x  \partial_x\bm{C} \bm{s} +  \kappa_y  \partial_y\bm{C} \bm{s},
\label{eq:Taylor1}
\end{equation}
where $ \partial_j\bm{C}$ is the derivative of $\bm{C}$ along $j\in\{x,y\}$, multiplied by the length scale $\Lambda$. Over one macro-pixel $m$ defined by one grid period, Eq.~\eqref{eq:Taylor1} can be written as a matrix system, assuming that the beam state is uniform over the macro-pixel extent
\begin{align}
\bm{I_m} \approx  \bm{A_m} \bm{\hat{s}},
\label{eq:Taylor2}
\end{align}
where $\bm{A_m} = (\bm{C},\partial_x\bm{C},\partial_y\bm{C})$ is the 12-column transfer function matrix, and $\bm{\hat{s}} = \left(\bm{s}^\mathsf{T}, \bm{s_x}^\mathsf{T}, \bm{s_y}^\mathsf{T}\right)^\mathsf{T}$ the vector describing the corresponding beam state. For a spatially coherent beam, Eq.~\eqref{eq:Taylor2} must be solved subject to $\bm{s_x}=\kappa_x\bm{s}$ and $\bm{s_y}=\kappa_y\bm{s}$, thus imposing that $\bm{s}$, $\bm{s_x}$ and $\bm{s_y}$ are colinear. In order to allow the simultaneous determination of the vectors $\bm{s}$ and $\bm{\kappa}$, Eq.~\eqref{eq:Taylor2} must always be invertible, for all $\bm{s}\neq \bm{0}$. We thus optimized the birefringent Hartmann mask for any possible beam state. To achieve so in the most general case, we considered minimizing the Cramer-Rao lower bound of the $12$-dimensional system (Eq.~\eqref{eq:Taylor2}), based on the calculation of the Fisher information matrix~\cite{kay1993cramer}. Assuming a normal noise distribution with uniform variance over all the camera pixels, the Fisher information matrix can be shown to be proportional to $\bm{F}\propto \bm{A_m}^\mathsf{T}\bm{A_m}$ (Supp.~\ref{app:fisher}). We chose to optimize the birefringent mask by minimizing the highest Cramer-Rao lower bound, corresponding to the least well-conditioned eigenstate. This is obtained by minimizing the condition number of $\bm{A_m}$, namely the ratio $\lambda_{1}/\lambda_{12}$, where $\lambda_1$ and $\lambda_{12}$ are the largest and the smallest eigenvalues of $\bm{A_m}$, respectively. In practice we conceived the mask as a birefringent square grating with a unit cell composed of $2\times 2$ domains of uniform slow-axis orientations (See Fig.~\ref{fig:principle}b). Due to technological limitations in the fabrication process of the liquid-crystal (LC) mask, the retardance is uniform all over the mask. A genetic algorithm was then used to jointly optimize the parameters of the grating: the retardance, the $4$ slow-axis angles in the unit cell, and the propagation distance $d$ between the mask and the camera plane. This optimization process converges towards the pattern shown in Fig.~\ref{fig:principle}b exhibiting slow-axis angles $\theta$ with spacings of $\pi/3$ and a retardance $\varphi =1.16\pi/2~{\rm rad}$. 
Several optimal distances $d$ to the camera are possible: $d\in\left\{0.22,0.28,0.72,0.78\right\}\Lambda_T$, where $d$ is expressed in units of the Talbot length $\Lambda_T = 2\Lambda^2/\lambda$. 
In the vicinity of the camera plane, the analytical expression of the cross-circular-polarized light-field converted by this mask is
\begin{multline}
E_\mp\propto 1\pm i\sqrt{3}\tilde{\nu} \left(\sin x-\sin y\right)\\
-\frac32\tilde{\nu}^2\left[\cos(x+y)-\cos(x-y)\right],
\label{eq:diff_orders}
\end{multline}
where $\tilde{\nu}=\frac{4}{\pi}\exp\left(-i\pi d/\Lambda_T\right)$ (Supp.~\ref{app:jones}). The field in Eq.~\eqref{eq:diff_orders} thus exhibits nine Fourier components arranged in a $3\times 3$ square (Fig.~\ref{fig:principle}c).
In Fig.~\ref{fig:principle}d, four typical experimental calibration intensity patterns are shown over two grating periods. Interestingly, the result of the birefringent-Hartmann-mask optimization is an array of focal points that closely resembles those utilized in conventional Shack-Hartmann wavefront sensing schemes. For the sake of illustration, an optional square-shaped aperture block (AB in Fig.~\ref{fig:principle}a and ~\ref{fig:principle}c) was deliberatly introduced to select, and exhibit the role of, the nine main orders at play. As discussed later, this spatial filter is not necessary in the following configuration where higher spatial frequencies are naturally filtered by the camera sampling at the Shannon-Nyquist criterion limit.

\begin{figure}[h!]
\centering
\fbox{\includegraphics[width=\linewidth]{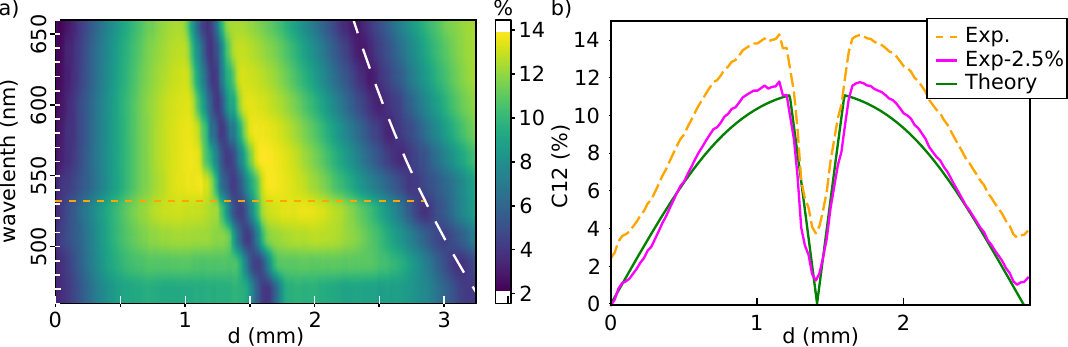}}
\caption{ {\bf Reciprocal condition number $\bm{C_{12}}$ of the transfer matrix.} Experimental values are shown in (a) as a function of the wavelength and the distance between the mask image and the camera $d$, demonstrating workability over a large spectral range. 
The white dashed line represent half the Talbot length $\Lambda_T$, the pattern being symmetrically repeated. A line profile is shown in (b) for $\lambda = 532~{\rm nm}$ ($\Lambda_T = 5.72~{\rm mm}$) and compared to the theoretical curve. The experimental curve exhibits a $\simeq 2.5\%$ extra offset.
} 
\label{fig:WD}
\end{figure}

A mask was manufactured (Thorlabs Inc., USA) according to the optimized design parameters given above and shown in Fig.~\ref{fig:principle}b. The reciprocal condition number $C_{12}=\lambda_{12}/\lambda_{1}$ of the corresponding transfer function matrix $\bm{A_m}$ is shown in Fig.~\ref{fig:WD}: $C_{12}$ is plotted as a function of the wavelength over a $200~{\rm nm}$ spectral range, and the distance $d$. The experiment was carried out using a fiber-coupled supercontinuum laser (ElectroVis-470, LEUKOS, France) combined to a computer-controled filter box
(Bebop, LEUKOS, France) providing bandwidths down to $5~{\rm nm}$. The line profile at the design wavelength ($\lambda = 532~{\rm nm}$) is shown in Fig.~\ref{fig:WD}b, demonstrating good agreement with numerical expectation up to a favorable experimental extra offset of $2.5\%$ that we attribute to fabrication imperfections. It must also be pointed out that these results demonstrate that the mask can effectively operate over a broad spectral range due to the achromatic nature of the phase imprinted onto the converted cross-polarized circular beam, which is solely determined by the geometric patterning of the birefringence axis angle.  

Since the polarimetric and wavefront information is fully carried by the square-arranged $3\times 3$ main diffracting orders generated by the mask, the power spectrum of the intensity at the camera -- given by the auto-correlation of this pattern -- contains $5\times 5$ spatial frequencies. Therefore, the minimal size of a macro-pixel is $5\times 5$ camera pixels for the pattern to be resolved (and the matrix $\bm{A_m}$ to be invertible). In this configuration, the aperture block (AB) shown in Fig.~\ref{fig:principle}a and~\ref{fig:principle}c is not necessary.

\paragraph{Image reconstruction.}
Eq.~\eqref{eq:Taylor2} must be solved over each period of the diffraction grating that represents one sampling point, both for the polarization state and the wavefront gradient. Numerically, we solved this equation considering a sliding window, using the Lucas-Kanade method weighted by a Gaussian window~\cite{Lucas_Kanade_81}. This method consists in assuming that the sought-for state $\bm{\hat{s}}$ (polarization state and wavefront gradients) is uniform over the considered Gaussian window width (the close-neighbor approximation). In practice, we used a Gaussian waist $w=2/3\Lambda$. 
To solve Eq.~\eqref{eq:eq_WFS} for distortions beyond the frame of validity of the first order Taylor expansion, we implemented the inversion of Eq.~\eqref{eq:Taylor1} in an iterative version, wherein after every inversion of Eq.~\eqref{eq:Taylor2}, the calibration intensity patterns are registered according to the displacement maps $\bm{\kappa}$ measured at the former step. We found that the algorithm typically converges after a few steps. As a result, image resolution was improved and wavefront measurements were made quantitative.

\paragraph{Results.}

\begin{figure}[!h]
\centering
\fbox{\includegraphics[width=\linewidth]{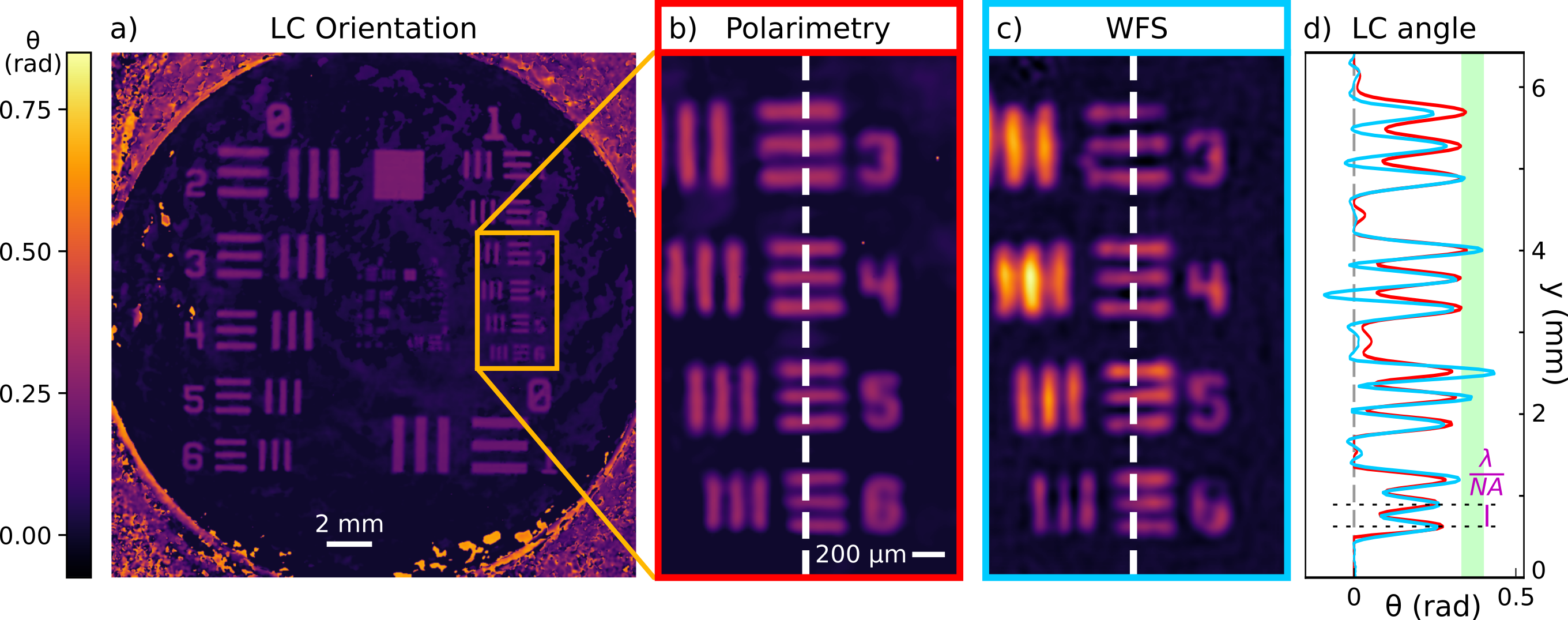}}
\caption{ {\bf Quantitative phase and polarimetric measurements of a birefringent USAF target.} (a) Full-field image of the orientation $\theta$ of the LC of the target obtained through polarimetric measurements. (b) Zoom of a region reaching the resolution limit of the instrument. (c) The orientation $\theta$ over the very same region is obtained from the wavefront of the cross circular-polarized contribution of the beam. (d) Line profiles of $\theta$ along the white dashed lines in (b) and (c). The green strip in (d) indicates the additional experimental measurement of $\theta$ in motifs estimated by rotating the sample between two crossed polarizers. The smallest resolved lines are very close to the expected resolution limit $\lambda / \mathrm{NA}$.
} 
\label{fig:USAF}
\end{figure}

\begin{figure*}[h!]
\centering
\fbox{\includegraphics[width=\linewidth]{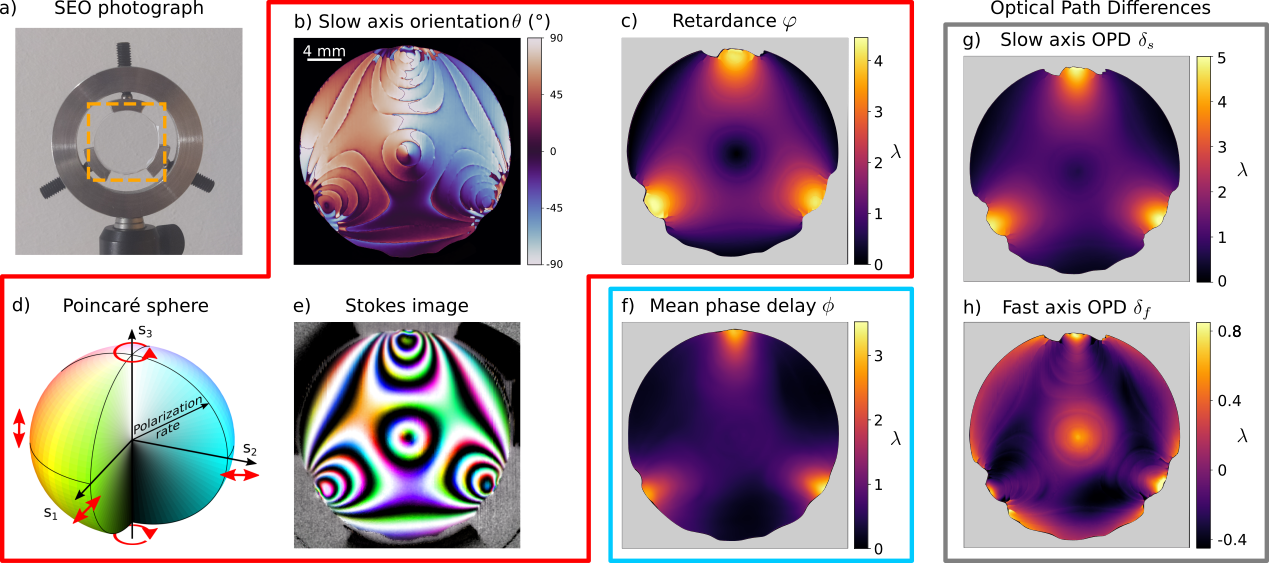}}
\caption{ {\bf Single-shot polarimetric (red box) and wavefront (blue box) imaging of an SEO.} (a) Regular photograph of the custom-made SEO. (b) slow axis orientation $\theta$ exhibiting the topolotical singularity and (c) retardance $\varphi$, both obtained from the polarimetric measurement. (d,e) The full-Stokes image of the SEO obtained from an incoming $\bm{\sigma_-}$ polarization state (e) together with its color coding of the polarization state represented on the Poincar{\'e} sphere (d). (f) The mean phase delay $\phi$ recovered by integration of the displacement map $\bm{\kappa}$. (g,h) Optical path differences of the slow (g) and fast (h) axes of the SEO obtained from the retardance $\varphi$ and the mean phase delay $\phi$.
} 
\label{fig:SEO}
\end{figure*}

Hereafter, samples are illuminated using a white LED filtered by a $532/10~{\rm nm}$ excitation filter. The beam was collimated using a $150~{\rm mm}$ focal length lens after passing through a $100~{\rm \mu m}$ pinhole, and polarized using a custom-made computer-controled polarization state generator.

To demonstrate the performance of the quantitative phase and polarimetric imager, a linear birefringent USAF target (R3L1S1B, Thorlabs Inc., USA) 
was imaged (Fig.~\ref{fig:USAF}a). This sample was chosen because it allows for a cross-validation of the accuracy of polarization and wavefront measurements. Namely, according to Eq.~\eqref{eq:bir-plate}, the orientation of the LC molecules $\theta(x,y)$ can be retrieved both from the polarization image reconstruction only (Fig.~\ref{fig:USAF}b) or from the wavefront image of the converted crossed circularly polarized component (Fig.~\ref{fig:USAF}c). To further validate these results, the LC molecules' orientation was also estimated by rotating the sample between two crossed polarizers (green strip in Fig.~\ref{fig:USAF}d). 

First, a ``ground truth'' $\theta$ image was obtained from polarization-only measurements by inverting a reduced version of Eq.~\ref{eq:Taylor2} including only ${\bm C}$. To improve the measurement accuracy, several acquisitions were performed with our device for several input polarization states. The LC orientation was then determined by fitting the measured Mueller matrix to a model of linear retarders using a Levenberg–Marquardt non-linear least square algorithm \cite{levenberg1944method}.
Second, both phase and polarimetric images were obtained from a single image acquisition obtained by illuminating the sample with a $\bm{\sigma_-}$ polarized beam. The $\bm{\sigma_+}$ converted beam thus carries a spatial phase $e^{-2i\theta}$ (Eq.~\eqref{eq:bir-plate}).
In our iterative algorithm, the dimension of $\bm{A_m}$ in Eq.~\eqref{eq:Taylor2} was reduced to $6$ to improve the numerical stability of the matrix inversion. $\bm{A_m}$ was defined as: 
$\bm{A_m} = (\bm{C}, \partial_x\bm{c_{+}}, \partial_y\bm{c_{+}})$, where $\bm{c_+}=(\bm{c_0}+\bm{c_3})/2$ is the normalized calibration intensity pattern associated with the circular polarization state $\bm{\sigma_+}$. 
This simplification is justified by the fact that in the present case, the retardance of the sample ($\varphi = 280~{\rm nm}$ according to the manufacturer) corresponds almost to that of a half-wave plate at our working wavelength ($\lambda = 532$~{\rm nm}) and thus almost fully converts the impinging circular polarization state into the cross-polarized state (less than $1\%$ left in the $\bm{\sigma_-}$ channel). The pattern gradients of other polarization states are thus not necessary to find the wavefront gradient of the $\bm{\sigma_-}$ beam.

Third, an additional experimental validation was performed by rotating crossed-polarizers and revealed that the LC on the patterns are rotated by $21 \pm 2^\circ$ relatively to the background ones (green strip in Fig.~\ref{fig:USAF}d). 

As a result, polarimetric measurement (Fig.~\ref{fig:USAF}b), wavefront measurements (Fig.~\ref{fig:USAF}c) and our third validating experiment exhibit good agreement with one another, as can be seen on the line profiles in Fig.~\ref{fig:USAF}d. The closest resolved lines in the sample are spaced by $\simeq 290~{\rm \mu m}$, very close to the estimated resolution limit of our imaging system $\lambda / \mathrm{NA} \simeq 270~{\rm \mu m}$. 

Next, a custom-built stress engineered-optic (SEO)~\cite{Brown_AO_07,Brown_OE_07} was considered as a sample (Fig.~\ref{fig:SEO}a). SEOs have been used for producing structured polarization beams~\cite{Beckley:10,Beckley:12}, for polarization measurements of sparse objects~\cite{Ramkhalawon2013, Zimmerman:16, Sivankutty:16} and as Fourier filters for dipole spread function engineering~\cite{Vella2019}, in particular for super-localization microscopy~\cite{Curcio2020a}. The SEO used here consists of a circular window of N-BK7 glass stressed at three points arranged in an equilateral triangle. This stress induces an anisotropic linear birefringence pattern exhibiting a $-1/2$-charged line dislocation at the center. A $\bm{\sigma_\pm}$ polarized beam thus yields a transmitted $\bm{\sigma_\mp}$ carrying orbital angular momentum. 
%
%
We imaged the SEO under $\bm{\sigma_-}$ polarized light beam illumination with the aim to characterize both the birefringence pattern as well as geometric aberrations (that are not visible by polarimetric imaging). 
Assuming that the light beam remains spatially coherent after passing through the sample, Eq.~\eqref{eq:Taylor2} is solved subject to the collinearity constraints $\bm{s_x}=\kappa_x\bm{s}$ and $\bm{s_y}=\kappa_y\bm{s}$.
We solved this system thanks to an iterative algorithm imposing hard constraints.
The algorithm consists in iteratively: (i) estimating $\bm{s}$ from the zeroth-order Taylor expansion of Eq.~\eqref{eq:eq_WFS}: $\bm{Cs} = \bm{I_m}$, (ii) synthesizing the corresponding calibration pattern $\bm{c_e}= \displaystyle{\sum_{j=0}^3} s_j\bm{c_j}$, (iii) solving $\bm{A_m}\bm{\hat{s}} = \bm{I_m}$ with $\bm{A_m}$, of rank three, defined as $\bm{A_m} = (\bm{c_e}, \partial_x\bm{c_e}, \partial_y\bm{c_e})$, and (iv) performing a non-rigid registration of $\bm{C}=(\bm{c_0},\bm{c_1},\bm{c_2},\bm{c_3})$. 
In a single acquisition step, we retrieve full Stokes information as well as wavefront distortions. Next, relying on Eq.~\eqref{eq:bir-plate}, the physical parameters of the SEO can be extracted: the slow axis orientation $\theta$ (Fig.~\ref{fig:SEO}b), the retardance $\varphi$ (Fig.~\ref{fig:SEO}c) and the average phase delay $\phi$ (Fig.~\ref{fig:SEO}f). The Stokes image is shown in Fig.~\ref{fig:SEO}e with a Hue-Saturation-Lightness color code (Fig.~\ref{fig:SEO}d). 
%
%
Considering an impinging $\bm{\sigma_-}$ circular polarization state on the SEO, from Eq.~\eqref{eq:bir-plate} we obtain:
\begin{align}
    \tan^2\left(\frac{\varphi}{2}\right) &= \frac{s_0+s_3}{s_0-s_3}, \label{eq:pola_phi} \\
     ie^{-2i\theta}\sin\varphi &= s_1+is_2,
    \label{eq:pola_theta}
\end{align}
so that $\varphi$ and $\theta$ can thus be computed up to a sign ambiguity accounting for the uncertain identification of the fast and slow axes when imaging at a single wavelength. For this reason, their respective images in Fig.~\ref{fig:SEO}b and Fig.~\ref{fig:SEO}c have been unwrapped: the sign was changed for every zero crossing of $\sin\varphi$, starting from the center where $\varphi$ is known to be zero for symmetry reasons.
The slight discontinuity lines in Fig.~\ref{fig:SEO}b are due to this unwrapping process. The slow axis orientation accounts for the refractive index increase along constraint lines.
Finally, the integration of the displacement map $\bm{\kappa}$ allows recovering the average optical path difference map $\phi$ imprinted to the beam (Fig.~\ref{fig:SEO}f). From $\phi$ and $\varphi$, the slow and fast OPDs, $\delta_s$ and $\delta_f$, respectively, can be plotted (Figs.~\ref{fig:SEO}g and~\ref{fig:SEO}h). The ability to measure not only the retardance of the sample but also the average OPD demonstrates the practical interest of coupling polarimetric and wavefront imaging modalities to access full optical characterization.

\paragraph{Conclusion and discussion.}
We demonstrated the possibility of extending the principle of multiwave lateral shearing interferometry to vector beams and to encode the full Stokes polarization state of a beam in addition to its wavefront. Since more information is multiplexed at the camera plane, more interfering waves are required and one sampling point requires more camera pixels ($5\times 5$ at least) so providing $400\times 400$ full-Stokes polarimetric and wavefront images from a single $4~{\rm Mpx}$ camera. 

Interestingly, the dimension of our optimized transfer function matrix  $\bm{A_m}$ in Eq.~\eqref{eq:Taylor2} is $12$. If relaxing the constraint about the colinearity of $\bm{s}$, $\bm{s_x}$ and $\bm{s_y}$, our polarimetric WFS instrument turns out to be sensitive to the polarization-resolved spatial coherence of the light source~\cite{Sanchez_Soto_14}. Indeed, considering a density of incoherent classical emitters $\rho(\bm{s},\bm{\kappa})$, the intensity at the camera sensor is $\bm{I} = \bm{C}\overline{\bm{s}}+ \partial_x\bm{C} \overline{\kappa_x\bm{s}}+ \partial_y\bm{C} \overline{\kappa_y\bm{s}}$ where the bar notation denotes averaging over the emitters distribution: $\overline{\kappa_j\bm{s}}=\iint \kappa_j\rho(\bm{s},\bm{\kappa}){\rm d}^4\bm{s}{\rm d}^2\bm{\kappa}$. 

Finally, the mask optimized in this manuscript is designed based on specific statistical assumptions regarding noise and wavefront distortion magnitudes but other masks can be equivalently designed for different cases following the same principles.


\paragraph{Acknowledgments.}
The authors acknowledge Sophie Brasselet and Pascal Berto for stimulating discussions and Dorian Bouchet for his remarks about the manuscript. This work was partly funded by the French Research National Agency (project PhaPIm) and the Future Investment Program of the French government (Stratex Universit{'e} Paris Cit{'e}). M.G. acknowledges support from Institut Universitaire de France and M.A.A acknowledges support from ANR-21-CE24-001401.


\bibliographystyle{ieeetr}
\bibliography{240604_polar}


\clearpage

\newpage

\beginsupplement
\setcounter{page}{1}

\twocolumn[{%
\centering
\bigskip
\huge {A Polarimetric Wavefront Imager:\\Supplementary Materials }\\[1em]
\large{Baptiste Blochet$^{1\ast}$, Gr{\'e}goire Lelu,$^{1}$ Miguel A. Alonso,$^{2}$ Marc Guillon$^{1,3,4\ast}$}\\[1em]
\noindent
\normalsize{$^{1}$ Saints-Pères Paris Institute for the Neurosciences, CNRS UMR 8003, Universit{\' e} Paris Cit{\' e},}\noindent\\
\normalsize{ 45 rue des Saints-Pères, Paris 75006, France,}\\
\normalsize{$^{2}$Aix Marseille Univ, CNRS, Centrale Med, Institut Fresnel, UMR 7249, 13397 Marseille, France}\\
\normalsize{$^{3}$Institut Langevin, ESPCI Paris, Universit\'e PSL, CNRS, Paris 75005, France}\\
\normalsize{$^{4}$Institut Universitaire de France, Paris, France}\\
\normalsize{$^\ast$E-mail:  baptiste.blochet@u-paris.fr - marc.guillon@u-paris.fr}\\[3em]
}]


\section{Transmission matrix of a linear birefringent plate, in the circular polarization basis.}
\label{app:linear_retarder_plate}

In the canonical basis, the Jones matrix of a linear retarder plate is
\begin{gather}
J_{plate}
=
\begin{bmatrix}
e^{i\delta_s} & 0 \\
0 & e^{i\delta_f}
\end{bmatrix}.
\end{gather}
In the x-y basis, the plate is rotated by an angle $\theta$ and the Jones matrix is obtained as
\begin{equation}
J(\theta) = R(\theta)JR(-\theta),
\end{equation}
with $R(\theta)$ the rotation matrix, given by 
\begin{gather}
R(\theta)
=
\begin{bmatrix}
\cos(\theta)& -\sin(\theta) \\
\sin(\theta)&\cos(\theta)
\end{bmatrix}.
\end{gather}
This leads to
\begin{gather}
J(\theta)
=
\begin{bmatrix}
\cos^2(\theta)e^{i\delta_s}+\sin^2(\theta)e^{i\delta_f}& \cos(\theta)\sin(\theta).(e^{i\delta_s}-e^{i\delta_f} )\\
\cos(\theta)\sin(\theta).(e^{i\delta_s}-e^{i\delta_f} ) &\sin^2(\theta)e^{i\delta_s}+\cos^2(\theta)e^{i\delta_f}
\end{bmatrix}.
\end{gather}
For a circular incident polarization $E_{\pm}$, the output field is given by Eq.~\eqref{eq:bir-plate}

\noindent\fbox{%
\parbox{\columnwidth}{%
\begin{align}
\bm{E}_{out,\pm} &= e^{i\phi}\left[ \cos\left(\frac{\varphi}{2}\right)\bm{\sigma_\pm} +i e^{\pm 2i\theta} \sin\left(\frac{\varphi}{2}\right)\bm{\sigma_\mp} \right].
\end{align}
}}

with $\phi = \frac{\delta_s+\delta_f}{2} = h\frac{n_s+n_f}{2} $ the mean phase delay, $\varphi = \delta_s-\delta_f = h(n_s-n_f)$ the retardance and $\theta$ the slow axis orientation with respect to the $x$-axis. 

\section{Wavefront sensor equation for vector beams.}
\label{app:WFS_equation}

We consider here a spatially coherent incident electric field $\bm{E_{in}} = E_{in}\bm{e} = A\exp(i\phi)\bm{e}$, with amplitude $A$, spatial phase $\phi$ and polarization state $\bm{e}$. This field is transmitted through the polarization-modulating Hartmann mask defined by its $2\times 2$ complex transfer matrix function $J$. 
The output field $\bm{E_{out}}$ at the camera plane can be obtained by solving to the paraxial Helmholtz equation:
\begin{align}
2ik_0\frac{\partial \bm{E_{out}}}{\partial z} = -\nabla_\perp^2 (J\bm{E_{in}}).
\end{align}
Expressing  $\bm{E_{in}}= A\exp(i\phi)\bm{e}= E_{in} \bm{e}$, and substructuring, we find
\begin{align}
2ik_0\frac{\partial \bm{E_{out}}}{\partial z} = - [\nabla_\perp^2 (J\bm{e})] E_{in}- J\bm{e} \nabla_\perp^2 E_{in} - 2 \nabla_\perp (J\bm{e}) \cdot\nabla_\perp E_{in}.
\end{align}
In this equation $J\bm{e}$ is interpreted as the output field $\bm{E_{c,\bm{e}}}$ for a calibration beam having polarization state $\bm{e}$, flat wavefront and unit amplitude. Moreover, we have $-\nabla_\perp^2 E_{in} = 2ik_0\frac{\partial E_{in}}{\partial z}$ which is the paraxial Helmholtz equation for the impinging beam propagating in free space if no mask were present. \emph{In fine}, we get
\begin{align}
\frac{\partial \bm{E_{out}}}{\partial z} =\frac{\partial }{\partial z}(E_{in}\bm{E_{c,\bm{e}}}) - \frac{1}{ik_0}\nabla_\perp E_{in}\cdot\nabla_\perp \bm{E_{c,\bm{e}}} 
\end{align}
which, after integration along $z$, yields
\begin{align}
\bm{E_{out}} = E_{in}\bm{E_{c,\bm{e}}} + \frac{iz}{k_0}\nabla_\perp E_{in}\cdot \nabla_\perp \bm{E_{c,\bm{e}}}
\label{eq:WFS_eq_Helmholtz_field}
\end{align}
which can be interpreted as a first order Taylor expansion
\begin{align}
\bm{E_{out}} &= E_{in}\bm{E_{c,\bm{e}}}\left[\bm{r}+\frac{iz}{k_0}\nabla_\perp \ln (E_{in}) \right]\nonumber\\
&= Ae^{i\phi} \bm{E_{c,\bm{e}}}\left[\bm{r}-z\frac{\nabla_\perp \phi}{k_0}+iz\frac{\nabla_\perp \ln (A)}{k_0}\right].
\end{align}
Ignoring amplitude modulations, we get the wavefront sensor equation for the field:

\noindent\fbox{%
\parbox{\columnwidth}{%
\begin{equation}
\bm{E_{out}} = Ae^{i\phi} \bm{E_{c,\bm{e}}}\left(\bm{r}-z\frac{\nabla_\perp \phi}{k_0}\right)
\label{eq:WFS_eq_Helmholtz_field_2}
\end{equation}
}
}
The intensity at the camera plane can obtained by directly squaring Eq.~\eqref{eq:WFS_eq_Helmholtz_field_2} if ignoring the effect of intensity gradient. However, in order to underline its contribution, we rather square Eq.~\eqref{eq:WFS_eq_Helmholtz_field}:
\begin{align}
I_{out} &= I_{in}\left( {\bm{c_e}} - z\frac{\nabla_\perp\phi }{k_0} \cdot \nabla_\perp {\bm{c_e}} \right)+ \frac{z}{k_0} \left( \nabla_\perp I_{in}\cdot\nabla_\perp\phi_{\bm{c},\bm{e}} \right) {\bm{c_e}}\nonumber \\
&= I_{in} {\bm{c_e}} \left(\bm{r}- z\frac{\nabla_\perp\phi }{k_0}\right) + \frac{z}{k_0} \left( \nabla_\perp I_{in}\cdot\nabla_\perp\phi_{\bm{c},\bm{e}} \right) {\bm{c_e}}
\label{eq:WFS_eq_Helmholtz_intensity}
\end{align}
where ${\bm{c_e}} = |{\bm{E_{c,e}}}|^2$ is the calibration intensity for polarization state ${\bm e}$ and $\phi_{\bm{c},\bm{e}}$ is the phase of the calibration field at the camera: $\phi_{\bm{c},\bm{e}}=\Arg\left(\bm{E_{\bm{c},\bm{e}}}\right)$. We can see that the effect of amplitude gradients is related to $\nabla_\perp\phi_{\bm{c},\bm{e}}$. If ignoring this contribution, we finally get:
\noindent\fbox{%
\parbox{\columnwidth}{%
\begin{equation}
I_{out} = I_{in} {\bm{c_e}} \left(\bm{r}- z\frac{\nabla_\perp\phi }{k_0}\right)
\end{equation}
}
}
which can also be written in the form of Eq.\eqref{eq:eq_WFS} in the main text by noting that $\bm{c_e}=\bm{C}\bm{s}$.

\section{Fisher information and Cramer-Rao lower bound}
\label{app:fisher}

The Fisher information matrix of the polarimetric wavefront sensor is derived in this section for basing our numerical optimization of the birefringent grating. This calculation demands assuming a distribution function for the noise. 
The Fisher information matrix is defined as
\begin{equation}
{\cal F} (\bm{x})= -{\rm E}\left\{\left.  \frac{\partial^2}{\partial \bm{x}^2} \ln \left[ P\left(\bm{y} |\bm{x}\right) \right] \right|  \bm{x}\right\},
\end{equation}
where $P\left(\bm{y}|\bm{x}\right)$ is the probability density function of the measurement vector $\bm{y} = \left(n_1, \ldots, n_M\right)_{j\in[1,M]}$ on a given macro-pixel containing $M$ pixels, parametrized by $\bm{x}$, the beam state vector. 
%
Assuming a normal distribution, $P\left(\bm{y};\bm{x}\right)$ can be written as
\begin{equation}
P(\bm{y}|\bm{x}) = \frac{1}{ (2\pi)^{M/2} |\bm{\Sigma}|^{1/2}} \exp\left[ -\frac{1}{2}(\bm{y}-\bm{\hat y})^\mathsf{T}\bm{\Sigma}^{-1} (\bm{y}-\bm{\hat y}) \right] \nonumber
\end{equation}
where $\bm{\Sigma}$ is the covariance matrix of the noise, $|\bm{\Sigma}|$ is its determinant and $\bm{\hat y}$ the expected value:
\begin{equation}
\bm{\hat y} = \bm{A}\bm{x}.
\end{equation}
Assuming a uniform diagonal covariance matrix ($\Sigma = \sigma^2 {\rm I}$), we get a simple expression:
\begin{equation}
{\cal F} (\bm{x})= \frac{ 1 }{\sigma^2} \bm{A}^\mathsf{T}\bm{A}.
\end{equation}
Under this assumption about noise statistics, the Fisher information matrix is just the covariance matrix of the transfer function matrix $\bm{A}$, up to a constant scaling factor. Provided that our estimator $\bm{\hat{x}} =\bm{A}^{-1}I$ is unbiased, minimizing the Cramer-Rao lower bound corresponds to maximizing the smallest eigenvalue of $\bm{A}^\mathsf{T}\bm{A}$. The optimization of our system thus requires to get the condition number of the transfer function matrix $\bm{A}$ as close to unit as possible.

\section{Derivation of the Jones matrix of the mask}
\label{app:jones}

In the circular polarization basis, the Jones matrix for a wave retarder with linear eigenpolarizations can be written as \cite{Ainola:01,Vella:18}
\begin{align}
J_0=\Sigma_0\cos\left(\frac{\varphi}2\right)-i \left(q_1\Sigma_1+q_2\Sigma_2\right)\sin\left(\frac{\varphi}2\right),
\end{align}
where $q_1$ and $q_2$ are the two first Stokes parameters of the fast eigenpolarization (the third vanishing for a birefringent medium), and where we ignored a global phase factor $\exp(i\phi)$. Here we used the Pauli matrices (labeled for the circular polarization basis), defined as
\begin{align}
\Sigma_0 &=\left(\begin{array}{cc}1&0\\0&1\end{array}\right),\,\,\Sigma_1=\left(\begin{array}{cc}0&1\\1&0\end{array}\right)\\\nonumber
\Sigma_2 &=\left(\begin{array}{cc}0&-i\\i&0\end{array}\right),\,\,\Sigma_3=\left(\begin{array}{cc}1&0\\0&-1\end{array}\right).\nonumber
\end{align}

The distribution of the parameters $q_n$ over the pixels of the unit cell is simple:
\begin{subequations}
\begin{align}
q_1&=\begin{array}{|c|c|}\hline 1 & -1/2 \\ \hline -1/2&1\\ \hline \end{array}=\frac14\,\begin{array}{|c|c|}\hline 1 &1 \\ \hline 1&1\\ \hline \end{array}+\frac34\,\begin{array}{|c|c|}\hline 1 &-1 \\ \hline -1&1\\ \hline \end{array}\, ,\\
q_2&=\frac{\sqrt{3}}2\,\begin{array}{|c|c|}\hline 0 & 1 \\ \hline -1&0\\ \hline \end{array}\, .
\end{align}
\end{subequations}
We now approximate these periodic pixel coefficient distributions by low order Fourier series expansions. Let the variables $x$ and $y$ be normalized so that each pixel corresponds to an interval between two consecutive integer multiples of $\pi$. The resulting expansions are then
\begin{subequations}
\begin{align}
q_1&\approx\frac14+\frac34\nu^2 \sin x\sin y=\frac14-\frac38\nu^2[\cos(x+y)-\cos(x-y)],\nonumber\\
q_2&\approx-\frac{\sqrt{3}}4 \nu (\sin x-\sin y),
\end{align}
\end{subequations}
where $\nu$ is a parameter that depends on the fill factor of the pixels. For a pixel that fills the array, $\nu=4/\pi$.

We now add a short propagation distance. Given the spatial frequencies of these different components, they will accumulate different phases upon propagation. If we factor out the carrier phase of the constant terms, we find that this propagation corresponds simply to replacing $\nu$ with $\tilde{\nu}=\nu\exp(-i\pi d/\Lambda_T)$:
\begin{align}
J&\approx
\Sigma_0\,\cos\left(\frac{\varphi}2\right)-i \left(\tilde{q}_1\Sigma_1+\tilde{q}_2\Sigma_2\right)\sin\left(\frac{\varphi}2\right),
\end{align}
where we used the abbreviations $\tilde{q}_n$ for $q_n$ with $\nu$ replaced with $\tilde{\nu}$. The Jones matrix of the mask can thus be expressed as
\begin{align}
J = \left(\begin{array}{cc} \cos\left(\varphi/2\right) & 
\sin\left(\varphi/2\right)\alpha \\ \sin\left(\varphi/2\right) \alpha^\ast & \cos\left(\varphi/2\right)\end{array}\right)
\end{align}
where
\begin{align}
\alpha = \frac{i}{4}\left\{ 1\pm i\sqrt{3}\tilde{\nu}(\sin x-\sin y) +3 \tilde{\nu}^2 \sin x\sin y\right\}
\end{align}
(which yields the Eq.~\eqref{eq:diff_orders} of the main manuscript).

For an arbitrary partially polarized plane wave in the $z$ direction, the intensity can be written as $I(\boldsymbol{r})=\displaystyle{\sum_{n=0}^3}S_nT_n(\boldsymbol{r})$, where $S_n$ are the Stokes parameters of the incident wave and 
\begin{align}
T_n=\frac12{\rm Tr}\left[J\sigma_nJ^\dagger\right].
\end{align}
From this we can find
\begin{subequations}
\begin{align}
T_0&=\frac{1+\cos\varphi}2+\frac{1-\cos\varphi}2(|\tilde{q}_1|^2+|\tilde{q}_2|^2),\\
T_1&=\sin\varphi\,{\rm Im}(\tilde{q}_1),\\
T_2&=\sin\varphi\,{\rm Im}(\tilde{q}_2),\\
T_3&=(1-\cos\varphi){\rm Im}(\tilde{q}_2^*\tilde{q}_1).
\end{align}
\end{subequations}






\end{document}